# Ideal Relative Flow Distribution on Directed Network

Kardi TEKNOMO [a]

[a] *Associate Professor Ateneo Pedestrian and Traffic Computing Laboratory, Ateneo de Manila University, Quezon City, 1108, Philippines*
  *E-mail: teknomo@gmail.com*

**Abstract**: In this paper we propose a new concept to prioritize the importance of a link in a directed network graph based on an ideal flow distribution. An ideal flow is the infinite limit of relative aggregated count of random walk agents' trajectories on a network graph distributed over space and time. The standard ideal flow, which is uniformly distributed flow over space and time, maximize the entropy for the utilization of a network. We show that the simulated trajectories of random walk agents would form an ideal relative flow distribution is converged to stationary values. This implies that ideal flow matrix depends only on the network structure. Ideal flow matrix is invariant to scalar multiplication and remarkably it is always premagic. Demonstration of ideal flow to the real world network was fitted into Sioux Falls transportation network.

## 1. INTRODUCTION

Network flow studies has been the foundation of transportation network analysis. In transportation theory, network analysis can be categorized into two main parts: Network Structure and Network Utilization. Related to transportation field, network structure discusses connectivity, topology and path (including shortest and longest paths) on the network. Network utilization, on the other hand, are related to the flow (such as traffic assignment, maximum flow, etc.).

On the surface, network structure seems rather simple. We have *nodes* (or sometimes called *vertices*) to represent intersections and *links* (or sometimes called *directed edges*) to represent the road mid-blocks. Nodes also represent the source or the origin of the journey (such as home or factory) and the sink or destination of the journey (such as office or shop). Having only the network structure does not give us clue about how the network is used. Galloway and Thacker (2007) stated that graph theory is lacking of several important aspects that make the network itself. Agents that utilize the network and the dynamic aspect of the agent are not discussed in the network structure theory. This is where *network utilization* fits in.

The simplest network utilization is in the form of *counting* the number of agents (i.e. cars, trucks, busses, trains or pedestrians) that passes a link or a node. The aggregation of the counting of the number of agents over time is called *flow*. The flow on a specific link (i.e., mid-block) or node (i.e., intersection) is studied based on counting the agents (i.e., vehicles, pedestrian) that passed it. Recent technology on *tracking* is catching up, and the product of tracking an agent's position is a set of trajectories. With improvements in technology, it is now possible to collect actual individual trajectories on a network using existing tracking devices. In contrast to counting, tracking trajectories technologies do not require the observers to be in static locations. The trajectory of an agent is associated with agent identification number and its location data at certain predetermined time interval.

Counting the trajectories of agents over a network structure and put them into a matrix form produces a flow matrix. A *flow matrix* is a square matrix where each element gives the number of trajectories that have passed through the corresponding directed edge (link). Each row (or column) of a flow matrix corresponds to a node in the network. Knowing the flow matrix of the actual network gives the transportation engineers and planners important knowledge on how to manage and control the traffic flow.

In this article, we propose a new perspective to view a flow matrix based on ideal conditions. This proposal is an indirect method to obtain another type of flow matrix. Instead of improving the existing traffic assignment methods or getting OD flow from traffic count, our proposal is to start from the ideal conditions (hence, we call it as *ideal flow*). One technique to find an innovative solution in TRIZ (Gadd (2011)) is coming from the ideal solution and bring it down to practical solution. Though we know that due to the nature of road network and user behaviour, the real traffic flows would never be uniformly distributed, the ideal solution would give use the guidance and direction of the best solution in which the traffic flow should be aimed. Using this approach, we could measure how efficient the current traffic flow compared to the ideal condition. The ideal flow matrix can be used as the guidance to manage the actual traffic flow such that the actual flow matrix will be transformed as close as possible to the ideal flow matrix. Through network optimization, it is also possible to estimate the actual flow based on the changing of the probabilistic parameters of the ideal condition. The term "ideal flow" was chosen because the method is based on steady state solution. Most of practical queuing formulas that are used in transportation network are also based steady state solution.

In this paper, we attempted to introduce another point of view from the traditional direct or indirect approaches to obtain flow matrix. Our new approach is to use transition probability matrix of random walk instead of an OD matrix. If one can view that in each intersection in a network, the flow to any direction can be viewed as probability distribution, then aggregating these flow probability distribution in each node, yield us a transition probability matrix that is also a stochastic matrix.

The purpose of this article is to present a novel network model to conceptualize the prioritization of the importance of a link in a directed network graph based on a relative traffic flow distribution. The tool is based on multi-agent random walk simulation on a network and linear algebra. The significances of this paper are as follow:

- Normalized flow based on trajectories of random walk on directed network converged to stationary distribution based on Markov Chain
- Relationship between random walk on network with the principle of maximum entropy
- Modeling of ideal flow by Monte Carlo method of random walk and hand computation (for small network) and linear algebra formulas.
- Application of ideal traffic flow for prioritization of edges using dynamic graph experiments
- We show the properties of ideal flow matrix which is premagic and invariant of positive scalar multiple.

## 2. RELATED WORKS

There are two traditional ways to obtain flow matrix, direct and indirect approaches. Given a road network, the flow matrix of the network can be obtained directly if we can count the number of agents (i.e. vehicles) that passed *each* of the links on the network for a specific time interval (i.e. a day). Unfortunately, such counting is usually prohibited by cost of the survey because ideally we need to count the number of agents for *all links* at the same time interval. Another newer direct approach is to install tracking devices (i.e. GPS) on the vehicles and gather the trajectories of vehicles as sampling on the road network as suggested by Yuan *et al* (2013).

Indirect method is utilized by knowing the demand of flow in term of origin destination (OD) of each agent and generate the agents trajectories through traffic assignment methods in computer as explained in detail in Bell and Iida (1997), Ortúzar and Willumsen (2001). Another indirect approach is to estimate the OD from a few traffic count on the actual road links [Bera and Rao (2011), Cassetta et al (1993)]. Still another indirect approach is to utilized mathematical relationship between generalized origin destination matrix and flow matrix as proposed by Teknomo and Fernandez (2014).

There is a research gap that the OD matrix is still the important element of both direct and indirect approaches above. A model to generate path sequence of the locations of an agent that moves based on random probability distribution is called random walk. Brownian motion is a subset of random walk where the probability distribution in continuous over time. Random walk on network graph has been studied extensively by Grimmett (2010), and Durrett (2007) to mention a few prominent studies. One particular interest within random walk theory is the stationary distribution of nodes. *Node stationary distribution* is the convergence results of the probability distribution to observe an agent at a node when the random walk is performed at the limit of infinite time step. In simulation of random walk, one could attempt to approximately mimic the infinite time step by setting a large number of time steps.

Network graph can generally be divided into two major categories: undirected graph and directed graph. The characteristics of stationary distributions of the two categories are different because undirected graph has the properties of symmetry on each link. One way to achieve node stationary distribution to form transition probability matrix, which has Markov property

$$\mathbf{T}_{ij} = \Pr\left[v_{t+1} = j | v_t = i\right] > 0, i \neq j \tag{1}$$

For undirected graph, Transition Probability Matrix is a symmetric matrix and equal to

$$\mathbf{T} = \Delta^{-1}\mathbf{A} \tag{2}$$

Where matrix $\Delta$ is a diagonal matrix of each node degree and $\mathbf{A}$ is adjacency matrix of the graph. The vector of stationary distribution of nodes is called Perron's vector denoted by $\pi$. Perron-Frobenius theorem states that the left Eigen vector of the transition matrix is associated to the maximal eigenvalue $\lambda = 1$ such that

$$\pi\mathbf{T} = 1\pi \tag{3}$$

The probability to observe an agent at a node in undirected graph does not depend on the global structure of the graph. The element of stationary distribution of undirected graph depends only the total number of its edges and the node degree, which is a local property of the node.

$$\pi_i = \frac{\deg(i)}{|E|} \tag{4}$$

For directed graph, Transition Probability Matrix is equal to

$$\mathbf{T}_{ij} = \begin{cases} \dfrac{1}{\deg_{out}(i)}, & i \to j \\ 0 & otherwise \end{cases} \quad (5)$$

Clearly the Transition Probability Matrix for undirected graph **T** is not symmetric and therefore it may have complex conjugated pair of eigenvalues. Node stationary distribution of general undirected graph is difficult to describe. When the undirected graph is irreducible and aperiodic, the random walk converges to a single stationary distribution of Perron vector $\boldsymbol{\pi}$.

Blanchard and Volchenkov (2011) stated that if directed graph G is periodic, the Transition Probability Matrix **T** can have more than one eigenvalues with absolute value one and the ratio between the maximum and the minimum elements of Perron's vector is

$$\frac{\max_i \boldsymbol{\pi}_i}{\min_i \boldsymbol{\pi}_i} \leq \max_i \deg_{out}(i)^{diam(G)} \quad (6)$$

Where diam(G) is diameter of irreducible digraph G. The ratio of Max/Min of Perron's vector is a normalization such that the sum of vector values would be 1.

Note that the stationary distribution on both directed and undirected graph on the literatures are on the node values (which is called the Perron's vector), not on the edge values (which is clearly form a matrix, not just a vector). In contrast, what we have found through simulation and reported in this paper is the stationary distribution of the flow (relative edge values) for undirected graph. The idea of finding flow on the network based on random walk is inspired by Franke (2007).

## 3. IDEAL FLOW

In this section, we present our main proposal of the theory of Ideal flow. We start with basic assumptions, definition, how to model and how to simulate (for medium size network) and how to compute manually (for small network) and linear algebra (for large network).

### 3.1 Basic Assumption

Majority of road transportation users are commuters. Every morning commuters are going to work or school almost the same time and every evening the same people go back home. When the commuters travel from home to their destinations and back, their choice of path are limited. They will not select to explore the city to the longest route just to go to their offices. In each intersection along the limited path of their choices, the probability distribution of each individual to turn left or right or to go straight are also limited to be almost the same over time. Aggregating the individual behavior of the probability distribution on each intersection would generate the collective probability distribution on each node in the network. Since the commuters' path of choices are finite, it is safe to assume that the collective probability distribution on each node in the network would remain stable over time. This is the main assumption of using Markov Chain such that we can obtain stationary distribution. Each node in the network is having a unique probability distribution on where to go to the next link from that node. The probability distribution on each node in the network is constant over time.

Hypothetically, we could actually conduct expensive surveys on each intersection and compute the probability distribution of it. However, as in any model, we can also set the probability distribution based on certain features. The most desirable features to be included in order to compute the probability distribution should be based on the inherent structure of the network itself.

We define the ideal condition for a traffic flow as the most distributed flow over time and space in the network. The reason we have traffic congestion is mostly because the flow is not well distributed over time and space. We have the term *peak hour traffic* to show that there is a highest demand for a network or a link over a short time period. It is often the case that the demand reaches the link capacity or even over the link capacity during the peak hour. One example of solution often suggested by the transport planners to cope with this problem is to shift the peak demand to be distributed over time through staggering work or school hours. In this idea, the purpose is to flatten the distribution of demand over time from distribution with one or two modes into ideal *temporal uniform distribution*. Similarly, if an origin destination flow can be distributed spatially over more independent alternative routes it would produce lower demand on each route than the one with less independent alternative routes. The principle of *spatial uniform distribution* is basically can be seen as infinite parallel servers in queuing theory (Gross *et al* (2008) shows that Poisson distribution that often used in Markovian queuing theory is identical to the joint density of the order statistics of uniform random variables). We imagine that the most distributed flow spatially and temporally may be indicated as the most efficient utilization of the network. This assumption would become the basic assumptions of ideal flow matrix. Thus, ideal flow happens when the flow is well distributed over time and space because it is the most efficient utilization of a network. We call the ideal flow that based on uniform distribution as *standard ideal flow*. We can show mathematically in later section of this paper that the efficiency of standard ideal flow can be computed as maximizing the information entropy on the network.

Uniform distribution assumption on a transportation network may be seen by some readers as unrealistic because road network structure are very nonhomogeneous. Freeways carry much higher capacity than arterials, streets have different numbers of lanes. Similarly pipeline network that deliver oils and water may also have main pipes and distributed pipes and local pipes. Having an ideal flow provides us with new understanding on the direction on what we are heading on new researches, not only extension of what have been known. To create something ideal, however, we must start from a certain set of parameters. The most extreme starting point is the most ideal condition which is uniform distribution (which implies the uniform capacity of each link) where the network entropy is maximized. The standard ideal flow which is based on uniform distribution can be generalized into more realistic generalized ideal flow (or simply call as *ideal flow*) to accommodate non-uniform link capacity. Mathematically speaking, such change are simply be done as another set of parameters of the same model. The network entropy can be computed in the same manner and obviously it is no longer maximized in the general cases.

Ideal flow is aggregation of infinite trajectories. In this infinite terms, simple driver's route choice should be seen in aggregate terms of probability distribution parameters rather than individual route choice. The convergence of the driver's route choice can be viewed as the generalized ideal flow. Similarly, because we are dealing with the ideal condition of infinite trajectories, there must be a paradigm shift that the current finite paradigm to use Origin Destination matrix is no longer applicable in the case of infinite trajectories.

**3.2 Ideal Flow Definition**

Consider an aperiodic irreducible simple directed network graph $G = (N, L)$ where $N$ is the set of nodes and $L$ is the set of directed links, occupied by agents of random walkers. In each time step each agent in the network moves from one node to the next permissible node. The count of agent's trajectories on each link is called *flow*. When a flow is normalized (i.e. either divided by the minimum flow or the maximum flow) in the network, it is called *relative flow*. We call the limit distribution of the relative flow for infinite simulation time as *ideal flow* because it represents the ideal condition when the agents are scattered over time and space. If the random walker agents select the possible out-links based on uniform probability distribution, we called the relative flow as *standard ideal flow*. If the decision of the random walkers to select the possible out-links is based on other probability distribution than uniform probability distribution, we call the relative flow as *generalized ideal flow* or simply *ideal flow*. An *ideal flow matrix* is a matrix $\mathbf{F} = [f_{ij}]$ where $f_{ij}$ is the ideal flow on a link from node $i$ to node $j$.

**Definition 1**: An **ideal flow** is the infinite limit of relative aggregated count of random walk agents' trajectories on a network graph distributed over space and time.

Ideal flow is based on the convergence of relative traffic flow distribution of multi-agent random walk simulation. Since random walk agents are basically moved based on a Markov Chain Transition matrix, ideal flow is basically the relative stationary distribution of flow on directed graph. *Ideal flow digraph* (IFD) is a useful tool to study the impact of structural change on the utilization of directed graph.

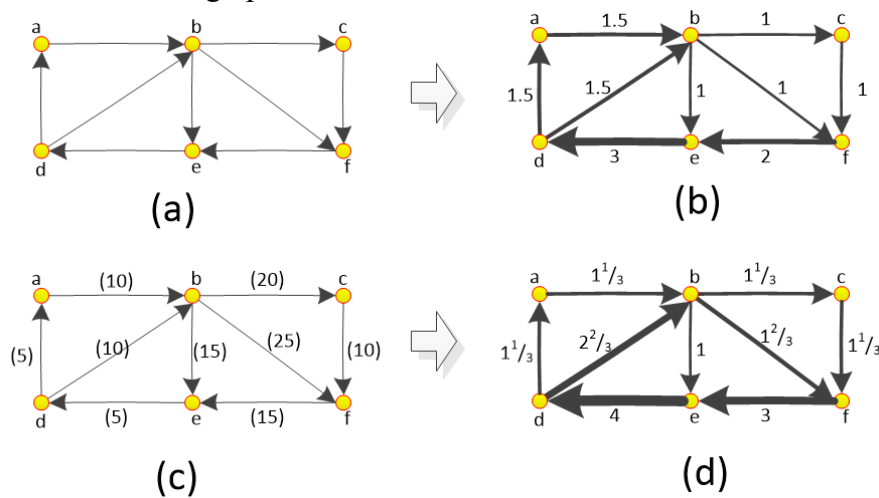

Figure 1. Example of directed graphs with their capacity ((a) and (c)). Figure (b) is the standard ideal flow of digraph (a) where the capacity are assumed to be equaled for all links. Figure (d) is the generalized ideal flow digraph (c). The percentage represent the relative capacity input between links that reflected in the output idea flows.

As an illustration in Figure 1, consider the following simple directed graph on the left and the ideal flow digraph on the right. The ideal flow number is also represented by the thickness weight of the arrows. The standard ideal flow (b) shows that the relative flow distribution with equal flow because the branches are under uniform distribution assumption. Implicitly, standard ideal flow assumes equal link capacity for all directed edges. The generalized ideal flow is computed based on the relative flow capacity as input (c). The output generalized ideal flow (d) will eventually reflect the relative capacity.

The readers shall not assume that the parameters of the distribution in Figure (1c) as merely for relative flow capacity only. This set of parameters is simply moldable generic parameters that can be used to model either the relative capacity or the actual flow. For instance, if the set of parameters represent the distribution of the current actual flow, then we actually model the ideal flow of the current flow. Assuming the probability distribution remain invariant, such model would be useful to perform what-if analysis to predict the ideal flow based on the structural changes of the network. Note that bidirectional links are the same as two directional links from the two nodes in reverse direction.

**3.3 Ideal Flow Modeling**

In this section we describe the modeling of the environment for ideal traffic assignment and modeling of the multi-agents. An ideal flow matrix is considered as the representation of an ideal of traffic flow distribution in a network. Instead of performing traffic assignment based on Wardop's principles of either social equilibrium or user equilibrium, we come up with a new model of ideal traffic assignment. *Ideal traffic assignment* is a random walk simulation of multi-agents to find the ideal relative traffic flow distribution.

A network under investigation must be a strongly connected directed graph (i.e. strongly connected or irreducible network) where there is path from any nodes to any other nodes. There is no self-loop on any node because self-loop flow will grow indefinitely. There is no source and no absorbing node such as only origin node or only destination node. Each node in the network can be viewed as origin and destination node. The network under investigation does not require any particular origin or destination matrix flow because any node in the network could be part of origin and destination of the agents.

We will now explain how to handle source and absorbing nodes through the introduction of a cloud node and dummy links. When a network under investigation is not strongly connected (reducible), one simplest strategy is to create a dummy node (symbolized by a cloud) to represent something outside the scope of the system as illustrated in Figure 2. The cloud node is connected only to the external nodes of the network. A dummy link can be provided from the cloud to each of the source nodes and from the absorbing nodes to the cloud. As an illustration, a network example below is representing weaving section of a road which is clearly not strongly connected. The right figure indicates the modeling approach to make the network become strongly connected.

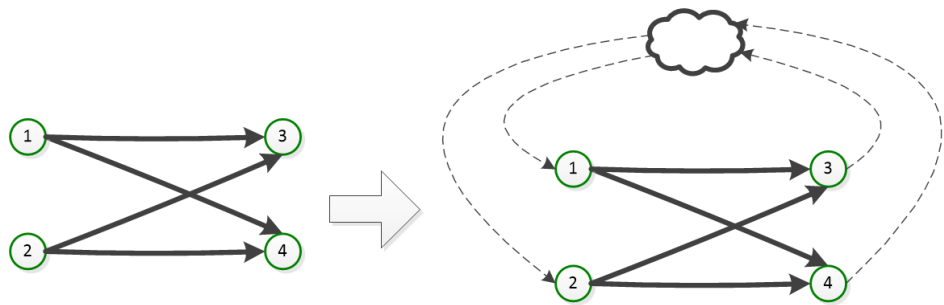

Figure 2. Example of adding dummy links from sink basins to source basin to transform the weakly connected network into strongly connected network for the computational purposes.

Computationally, ideal flow digraph can only be computed for strongly connected network. To compute ideal flow using computer, we need to connect the sink basins to the source basins with dummy links.

**3.4 Multi-Agent Simulation**

To create an ideal flow matrix, we cannot rely on the data from the actual traffic network which is clearly not ideal. Instead, we come up with the idea of random walk of finite multi-agents on network to generate the phenomena of spatial-temporal probability distribution of traffic flow. We imagine that we have N moving agents that will move around the edges of a given a directed graph network. At each intersection, each agent will decide random turn to any of the next edge to go based on prescribed probability distribution of all possible alternative edges. Each of the possible edges connected to the current edge has equal probability to be selected, which is based on the capacity and length of the next edge. Agents cannot move back immediately to the current edge (thus, U-turn must be modeled as more than one edge). One way to model the probability distribution as the set of input parameters is based on the relative capacity of all the links in the network.

Initially, the agents are located at random nodes. The source or the origin locations of the agents are bounded by the initial nodes but there are no destination nodes because the agents keep moving until the predetermined simulation time, T, is accomplished.

The trajectory movement of each agent on the network is recorded as dynamic trajectories data. From the entire trajectory data recorded, flow distribution matrix is computed and accumulated over time.

Counting the number of agent trajectories that passes an edge over the whole simulation time provide us with the actual count of edge flow. Putting these edge flows into a tabular form for all edges in the network with the same order as the adjacency matrix of the network yields a flow matrix $\mathbf{R}$.

When we take the ratio between the element of flow matrix and the sum of the total flow, we have relative flow matrix. Normalizing each non-zero elements of the relative flow matrix with the minimum (or maximum) of the non-zero elements of the relative flow matrix asymptotically gives us the ideal relative traffic flow matrix, or ideal flow matrix for short. The limit of the asymptote should be taken based on the product of large number of agents N and large number of simulation time T.

The following theorems are our major claims:

**Theorem 1**: When we increase the product of the number of agents and the simulation time to infinite term, the flow will change but the relative flow would be asymptotically converge to a certain value, which we called as *ideal flow*.

$$\mathbf{F} = \lim_{N \cdot T \to \infty} \frac{\mathbf{R}}{\min \mathbf{R}} \; if \; \mathbf{r}_{ij} \neq 0 \tag{7}$$

*Proof*:
Random walk of agents on a network can be modeled as Markov chain based on with Algorithm 1. Since the Markov Chain of aperiodic irreducible matrix is converged to Perron's vector, the flow out of the Perron's vector is also converged. QED

The following algorithms are used to obtain Markov transition matrix for standard ideal flow from adjacency matrix. The first step is to clean the data from self-loop. The second step is to compute the out degree. The third step is to set a diagonal matrix to represent the out degree of each node. The last step is to obtain the Markov transition matrix (based on equation (2)). The solution of the ideal flow can be obtained through solving equation (3).

---

**Algorithm-1: Markov Transition Matrix for Standard Ideal Flow**
**Input**:
- **A** is adjacency matrix size n of aperiodic irreducible network

**Output**:
- **T** is Markov transition matrix for standard ideal flow matrix

**Algorithm**:
1. $\mathbf{A} = (\mathbf{A} - diag(\mathbf{A})) > 0$ to delete the self-loop
2. Set vector of out degree $\mathbf{v} = \sum_j \mathbf{A}_{ij}$
3. Set diagonal matrix of each node degree $\Delta = diag(\mathbf{v})$
4. Set Markov transition matrix $\mathbf{T} = \Delta^{-1}\mathbf{A}$

---

Equation (7) shows astounding characteristic that an ideal flow matrix, which is relative traffic flow distribution matrix, is asymptotically invariant and inherently exist based on merely the network structure where the agents move using random walk. As the results of this ideal traffic assignment, the idea flow matrix is no longer depending on the actual origin-destination demand but it is depending only on the network structure alone.

**Numerical Example-1**:

As an illustration, suppose we have a random network with 5 nodes, each edge has equal capacity. The adjacency matrix is given by

$$\mathbf{A} = \begin{bmatrix} 0 & 1 & 1 & 1 & 1 \\ 0 & 0 & 1 & 1 & 1 \\ 0 & 1 & 0 & 1 & 1 \\ 1 & 0 & 0 & 0 & 1 \\ 0 & 0 & 0 & 1 & 0 \end{bmatrix}$$

We run 100 agents for 200 time steps and the simulation produces flow matrix as shown below. If we run until infinite time, the flow matrix would contain infinite flow, which is not so useful.

$$\mathbf{R}_{t=200} = \begin{bmatrix} 0 & 985 & 920 & 966 & 945 \\ 0 & 0 & 492 & 450 & 529 \\ 0 & 467 & 0 & 482 & 465 \\ 3806 & 0 & 0 & 0 & 3784 \\ 0 & 0 & 0 & 5709 & 0 \end{bmatrix}, \quad \mathbf{R}_{t=\infty} = \begin{bmatrix} 0 & \infty & \infty & \infty & \infty \\ 0 & 0 & \infty & \infty & \infty \\ 0 & \infty & 0 & \infty & \infty \\ \infty & 0 & 0 & 0 & \infty \\ 0 & 0 & 0 & \infty & 0 \end{bmatrix}$$

However, the magical properties happen if the increasing the number of agents and the simulation time is also followed by dividing the flow with the minimum values. Suddenly the asymptotic values approaches the ideal relative traffic flow distribution, which are integers in nature. Ideal flow matrix and the corresponding flow diagram where the weight represent the ideal edge flow is shown in Figure 3. More remarkably, the sum of rows is also always equal to the sum of columns, which we will call this property as *premagic*.

$$\mathbf{F} = \begin{bmatrix} 0 & 2 & 2 & 2 & 2 \\ 0 & 0 & 1 & 1 & 1 \\ 0 & 1 & 0 & 1 & 1 \\ 8 & 0 & 0 & 0 & 8 \\ 0 & 0 & 0 & 12 & 0 \end{bmatrix} \begin{matrix} 8 \\ 3 \\ 3 \\ 16 \\ 12 \end{matrix}$$
$$\phantom{\mathbf{F} = }\;\; 8 \;\; 3 \;\; 3 \;\; 16 \;\; 12 \;\; 42$$

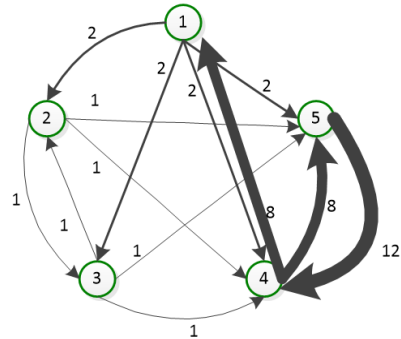

Figure 3. Illustration of ideal flows in Numerical Example-1:

### 3.5 Manual Computation of Standard Ideal Flow

As alternative to simulating N agents for T time steps, for relatively small network, we can do the computation of standard ideal flow manually (for small network). Algorithm-2 shows the steps of the manual computation.

| **Algorithm-2: Manual computation of Standard Ideal Flow** <br> **Input**: Directed Network Graph <br> **Output**: Ideal Flow on Directed Network Graph |
|---|
| 1. Set any node as origin. Set 100 flow into node origin <br> 2. When a node has only one in-edge and one out-edge, the same amount of flow continue from in-edge to out-edge <br> 3. When a node has more than one in-edge, sum all the flow from in-edges into the flow set in the node. <br> 4. When a node has more than one out-edge, distribute the flow equally among all out-edges. <br> 5. Nodes preserve the flow. All flow in into a node is equal to all flow out of the same node. <br> 6. At the end, we can get the ideal flow by dividing the flow with the minimum flow. |

### 3.5 Linear Algebra Computation of Ideal Flow

While the manual computation of ideal flow is good small network and the Multi-agent simulation of random walk approach is good for medium size network, we also need better approach for large network to produce ideal flow matrix. Random walk computation is rather slow to reach the asymptotical results. The manual computation has clear advantage of clear understanding of the process but it is very limited to small network. In this section, we device another approach to solve the ideal flow through linear algebra.

Suppose incidence matrix **B** (nodes on rows and edges on columns) can be standardized such that the edges are read from adjacency matrix row by row from left to right columns. Each node has an equation that all flows in is equal to all flows out. This equation is exactly formed in the incidence matrix. This is the balance equation or conservation of flow in each node.

Note that the signs in those balance equations in each node are the coefficient of incidence matrix. We can introduce edge vector (with size of the number of edges) **e** indicates the edge values. If p = sum of one entries adjacency matrix and n is the number of rows or number of columns in adjacency matrix which is equal to the number of nodes, the size of incidence matrix **B** would be n by p.

Thus, the edge equations on nodes are given by a simple homogenous system equation of

$$\mathbf{Be} = \mathbf{0} \qquad (8)$$

where vector **e** (size p by 1) indicates the edge values (the zero-vector has size n by 1). The null space of incidence matrix will produce the edge values **e**. However, this null space is not what we desire. The reason is because there are additional constraints that outflow of each node must be equal among all out-edges of the same node. To put these constraints into the equation, we find all the rows in the adjacency matrix that have sum of row larger than 1. Among these rows, for each two columns that has 1 entries we can for a constraint equation. Putting each of the constraint equations on the proper columns in the same order of edges as the incident matrix, we can add q rows of constraint equations, where q = p – n. Let **C** be q by p constraint matrix, we can augment incidence matrix and constraint to form

$$\mathbf{D} = \begin{bmatrix} \mathbf{B} \\ \cdots \\ \mathbf{C} \end{bmatrix} \qquad (9)$$

$$\mathbf{De} = \mathbf{0} \qquad (10)$$

Vector **e** (size *p* by 1) can be obtained as the null value of matrix **D**

$$\mathbf{e} = \text{null}(\mathbf{D}) \qquad (11)$$

Once vector **e** is determined through equation (11) above, we can verify equation (10) is correct.

How to transform matrix **B** into matrix **C** programmatically? In fact, we don't have to change matrix **B** into **C**. It is easier to obtain matrix **C** from adjacency matrix **A** directly because we can add constraint if we found sum of each row in adjacency matrix is larger than 1.

We can also separate the positive and negative values of the incidence matrix and we will call them $\mathbf{B}^+$ and $\mathbf{B}^-$ respectively.

$$\mathbf{B}^+ + \mathbf{B}^- = \mathbf{B} \qquad (12)$$

We can introduce node vector **n** (size n by 1, which is equal to the number of nodes) to indicates the flows accumulated in each node from all in-flows, or the source of all out-flows of a node. We can then obtain the following equations:

$$\mathbf{B}^+\mathbf{e} = \mathbf{n} \qquad (13)$$

$$\mathbf{B}^-\mathbf{e} = -\mathbf{n} \qquad (14)$$

The constraints can be obtained from the edges corresponds to the positive values of the incidence matrix. The flow values of all edges out of a node must be equal to each other and therefore we will have a set of constrained edge flows which is equal to the number of nodes. Let us call the constrained edge flow as vector $\mathbf{x}$. Our goal is to find the value of vector constrained edge flow $\mathbf{x}$. To do that, we manually set the edge vector $\mathbf{e}$ as function of constrained edge flow $\mathbf{x}$. Then, we solve the equation (8) to give the values of vector $\mathbf{x}$. Note that the null space of equation (8) is now constrained by the vector $\mathbf{x}$ and therefore the usual numerical method of null space computation through Singular Value Decomposition (SVD) will not work. Manual computation through substitution and elimination and generalize the values works only for small network. Setting the constraint matrix works to solve this problem.

Vector $\mathbf{e}$ indicates the flow ratio of edge values but it would be in a nicer format when we can obtain normalized of the ideal flow $\mathbf{f} = \dfrac{\mathbf{e}}{\min(\mathbf{e})}$ in such a way such that the minimum non-zero edge value is 1.

---

**Algorithm-3: Linear Algebra computation of Standard Ideal Flow**
**Input**: Directed Network Graph
**Output**: Ideal Flow on Directed Network Graph

1. Convert the adjacency matrix into incidence matrix $\mathbf{B}$
2. Separate the positive values of the incidence matrix $\mathbf{B}^+$
3. Set constraint matrix $\mathbf{C}$ based on $\mathbf{B}^+$
4. Set augmented matrix $\mathbf{D} = \begin{bmatrix} \mathbf{B} \\ \cdots \\ \mathbf{C} \end{bmatrix}$
5. Solve constrained edge flows $\mathbf{e}$ based on $\mathbf{De} = \mathbf{0}$
6. Flow ratio is normalized edge flow $\mathbf{f} = \dfrac{\mathbf{e}}{\min(\mathbf{e})}$

---

## 4. PROPERTIES OF IDEAL FLOW

In this section, we added several remarkable properties of ideal flow that are useful for interpretation and better understanding the ideal flow.

**Theorem 2**: Multiplying a positive constant to the ideal flow does not change the ideal flow.
$$\mathbf{F} \equiv \kappa \mathbf{F} \qquad (15)$$

*Proof*:
Since ideal flow is a relative flow distribution, we can always normalized it with either the maximum flow or the non-zero minimum flow. Thus, scalar multiplication to an ideal flow matrix does not change the ideal flow network because it can always be put back during normalization.

***Definition 2***: Premagic matrix $\mathbf{M}$ is defined as a square matrix that the vector sum of rows is equal to the vector sum of columns.

The sum of columns can be computed using matrix multiplication of row vector ones $\mathbf{j}^T$ with any matrix $\mathbf{A}$ to produce row vector of sum of each column $\mathbf{s}^T$. To get the vector sum of each rows $\mathbf{s}$, we multiply from the right.

***Definition 3***: sum of columns is defined as a row vector:
$$\mathbf{s}_1^T = \mathbf{j}^T \mathbf{A} \tag{16}$$

***Definition 4***: sum of rows is defined as a column vector:
$$\mathbf{s}_2 = \mathbf{A}\mathbf{j} \tag{17}$$

Thus, a premagic matrix is defined either by
- $\mathbf{s}_1 = \mathbf{s}_2$, or  (18)
- $\mathbf{s}_1^T = \mathbf{s}_2^T$

**Theorem-3:** Ideal flow matrix is always premagic matrix.
*Proof:*
Recalled that a matrix is premagic if and only of the sum of rows is equal to the sum of columns. The sum of rows represent the sum of out-flow of a node $\sum f_{out}$ and the sum of columns represent the sum of in-flow to a node $\sum f_{in}$. Node conservation of flow means the sum of inflow is equal to the sum of outflow $\sum f_{in} = \sum f_{out} = \pi$. Thus, if flow on a node is conserved, then the matrix must be premagic. QED

Thus, premagic matrix characterizes flow conservation on nodes.

**Theorem-4:** Ideal flow matrix can be obtained through Markov Chain.
*Proof:*
Based on equation (3), we can use Markov Chain to generalize the ideal flow formulation for any transition probability stochastic matrix $\mathbf{T}$ as shown in the following equations.

$$\boldsymbol{\pi} = \begin{bmatrix} \mathbf{T}^T - \mathbf{I} \\ \mathbf{j}^T \end{bmatrix} \backslash \begin{bmatrix} \mathbf{0} \\ \kappa \end{bmatrix} \tag{19}$$

Knowing the Perron's vector, the ideal flow matrix can be computed using Hadamard product of the stochastic matrix and the matrix of the Perron vector.

$$\mathbf{F} = \boldsymbol{\pi}\mathbf{j}^T \circ \mathbf{T} \tag{20}$$

QED.

## 5. MAXIMUM ENTROPY

The idea of standard ideal flow came from random walk on network. In this section, we will show the relationship between agent's random walk on a network graph and maximum entropy principle.

**Theorem-5**: Random walk on network with uniform random distribution maximize the network entropy.

*Proof*:

The probability distribution of the agents which best represents the current state of knowledge is the one with largest entropy. Subject to the constraints of the available information, we can seek the probability distribution which maximizes information entropy. Suppose in a node we have k number of edges to indicate the number of possible choices to go. The probability of to edge j would be $p_j$. Let $P = [p_1 \quad p_2 \quad \cdots \quad p_k]$. The entropy of the node would be

$$H = -\sum_{j=1}^{k} p_j \log_2 p_j \tag{21}$$

.

We would like to find the distribution that maximize the entropy of the node.

$$\max_{p_j} H = -\sum_{j=1}^{k} p_j \log_2 p_j \tag{22}$$

subject to

$$\sum_{j=1}^{k} p_j = 1$$

We use Lagrange multiplier to form Lagrangian

$$L(p_j) = H + \lambda g(p_j) \tag{23}$$

$$= -\sum_{j=1}^{k} p_j \log_2 p_j + \lambda \left( \sum_{j=1}^{k} p_j - 1 \right)$$

Setting the gradient of $L$ with respect to $p_j$ and equate it to zero, we have

$$\frac{\partial L(p_j)}{\partial p_j} = -\frac{1}{\ln 2}\left(1 + \ln(p_j)\right) + \lambda = 0 \tag{24}$$

Thus,

$$p_j = e^{\lambda \ln 2 - 1} = \text{constant} \tag{25}$$

Since the constraint equation $\sum_{j=1}^{k} p_j = 1$, we have

$$p_j = \frac{1}{k} \quad k = 1, 2, 3, \cdots \tag{26}$$

Equation (26) shows that the probability distribution of random walk on network which maximize the entropy is uniform distribution. QED.

This result is not surprising as it is also similar to the well-known Laplace's principle of indifference which is sometimes called the principle of insufficient reason in decision analysis in management science [Taha (2016)].

## 6. SIMULATION OF IDEAL FLOW TO DYNAMIC TRANSPORTATION NETWORK

The concept of Ideal Flow can be applied to provide us with tools to measure with some degree of accuracy what an ideal traffic flow in a network should be. To give an illustration on how we will use ideal flow, in this section we will apply the concept of in a simple connected network of 5 nodes. We do dynamic graph approach to add a single edge on each stage and compute the ideal flow distribution. Though we include edge capacity in our simple model, we can consider higher ideal flow means higher congestion level in that edge. Higher ideal flow on an edge also means this edge is more important than other edges. By the way of example, we will show that ideal flow can be used to know effect of potential congestion and how we can reduce the potential congestion.

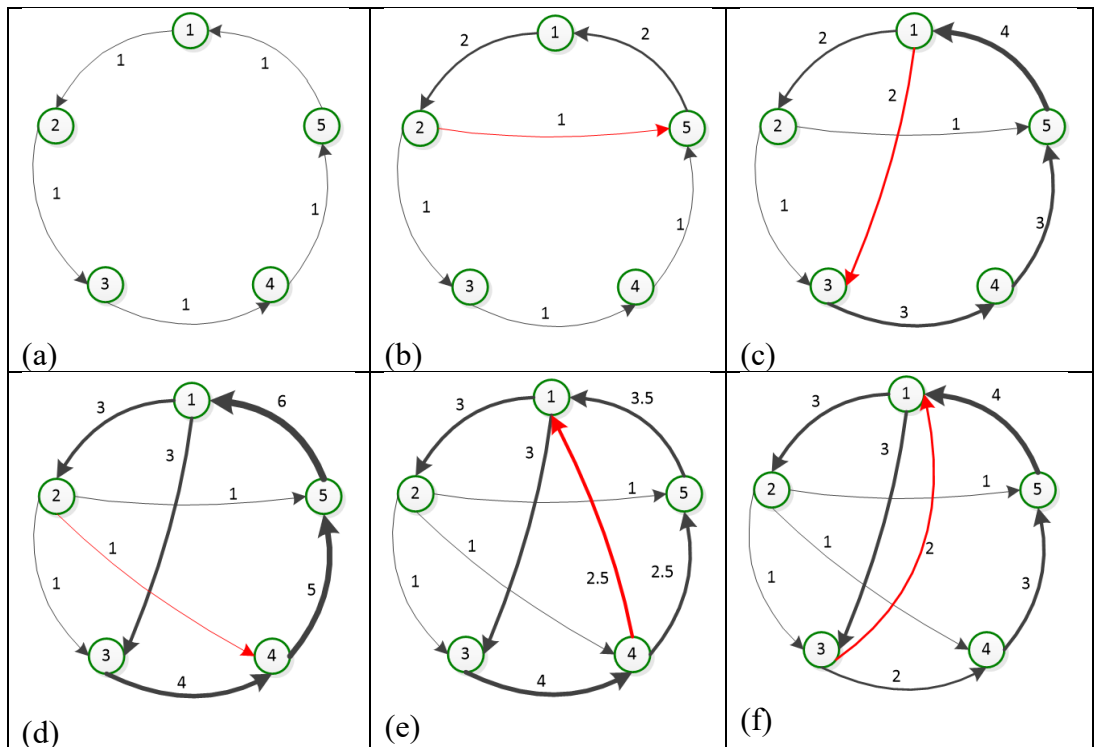

Figure 4. Example of Ideal Flow to determine relative congestion level using dynamic graph experiments.

We start with simple connected network where each node is connected with only one other node as illustrated in Figure 4 (a). The ideal flows in all edges are all one. Then we add edge 2-5 into the network (the red edge indicates the change). The ideal flow changes with edge 5-1 and 1-2 now become double ideal flows as shown in Figure 4 (b).

In Figure 4 (c), we add edge 1-3 and the maximum ideal flow become 4. Adding edges seems to increase the maximum ideal flow. If the same flow pass edge 2-3 (in which the ideal flow remains 1), edge 5-1 is now 4 times of edge 2-3.

In figure 4 (d) we add edge 2-4. Now the maximum ideal flow becomes 6. Edge 4-5 and 5-1 may become congested. We may want to reduce the congestion in those edges by adding edge that provide alternative route.

Now supposed we want to reduce the congestion by either adding edge 4-1 as illustrated in Figure 4 (e) or adding edge 3-1 as illustrated in Figure 4(f). Keeping the flow ratio of edge 2-3 to be 1 for all cases, it shows that adding edge 4-1 produces less congestion than adding edge 3-1.

Several conclusions can be derived from the simple experiment of dynamic graph above:
1. Adding a link is not always diffuse congestion. The fact that adding a link in a network may actually increase congestion seems a paradox but Braess Paradox is one example. In Figure 4 (b, c and d), additional edge actually increase the level of congestion.
2. Adding a link in certain link may cause congestion somewhere else far away from that added link. For instance, adding edge 2-4 in Figure 4(d) cause more congestion in edge 5-1 which is far away from the added edge 2-4. This unexpected results show the power of ideal flow approach in dealing with complexity.
3. A link can be added to divert the congestion by providing more direct alternative route and that link may contribute to reduce congestion. Figure 4 (e and f) show that congestion in Figure 4(d) with maximum ideal flow of 6 can be reduced to 3.5 (Figure 5e) or 4.0 (Figure 4f). Thus, we can conclude that adding edge 4-1 has more significant effect to reduce congestion that adding edge 3-1.

The simple illustration above have shown that adding edge on network may increase or decrease congestion level due to increase of importance level of the edge.

## 7. IDEAL FLOW OF SIOUX FALLS TRANSPORTATION NETWORK

To demonstrate the usefulness of ideal flow methods to model real world network, we use open network in GitHub called Sioux Falls Transportation Network as first published by LeBlanc (1975). The Sioux-Falls network has been used in many publications as a kind of standard test of traffic assignments because it is good for code debugging and also an opportunity to examine the data format. The network has 24 zones, 24 nodes, 76 links. A cloud node was added in the computation to model the source and absorbing nodes. The dummy links was added from each absorbing node to the cloud node and from the cloud node to each source node.

The link flow and OD matrix are given as the data. Note that the calibration does not need to use the OD matrix data. Since the flow are given, we can easily derived the transition probability matrix $\mathbf{T}$ from the flow data. One of the most important operation of ideal flow is the scaling operation as explained in equation (15). We then utilizing the equation (19) and (20) to search for the scaling parameter $\kappa$ that would minimize the sum of square error of the difference between the actual flow and the ideal flow. The scaling is applied to the whole network flow rather than to individual link flow. Our goal is to match the minimum sum-square of error between ideal flow and the actual flow in the data. The search for the scaling is shown in the top left figure below. The results of the calibration yield the scaling of $\kappa = 2{,}632{,}809.30$ and Mean Square Error of 562.05. The matched between the ideal flow and real world flow is shown in the top right figure 5 below.

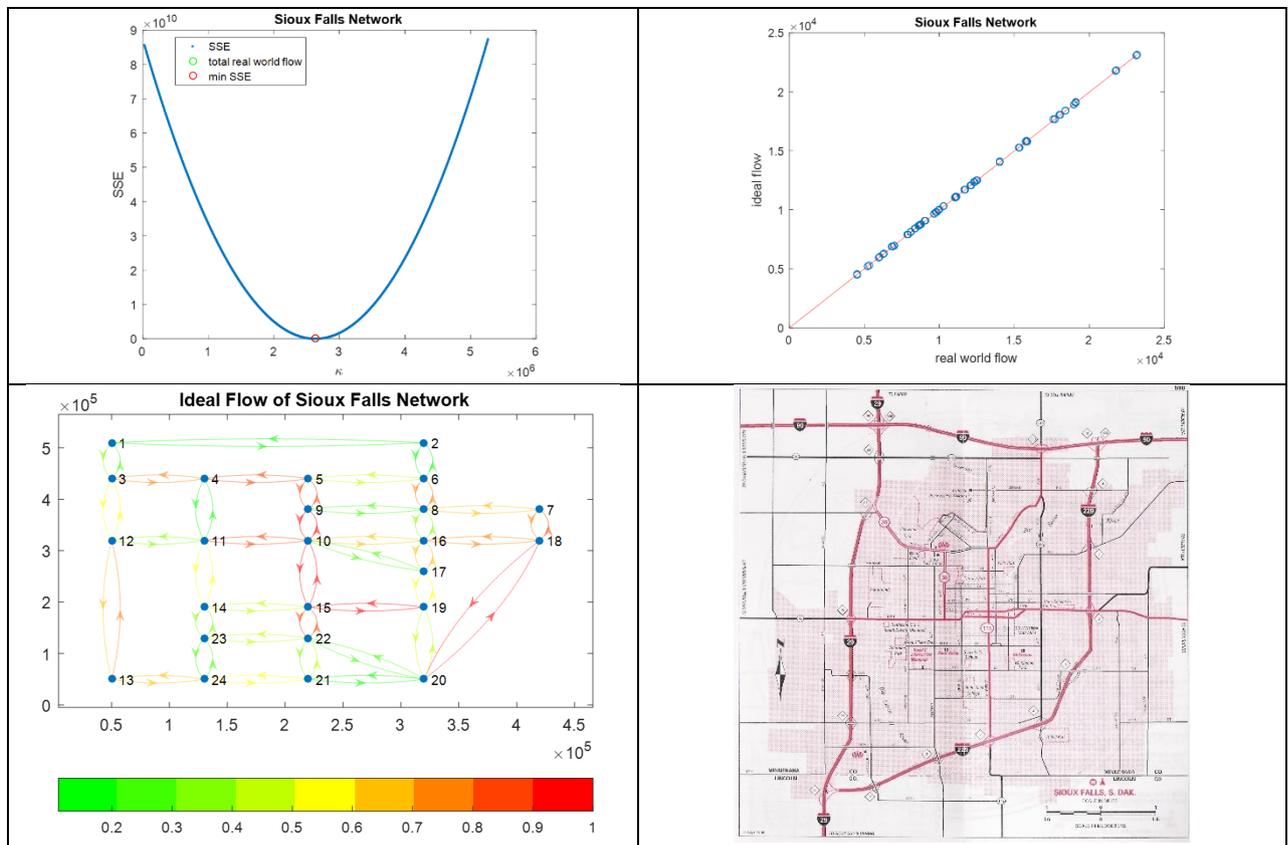

Figure 5. Applications of Ideal Flow to Real Transportation Network.

## 8. CONCLUSIONS AND RECOMMENDATIONS

Based on the analysis and discussion above, we may derive several conclusions:
1. The convergence of ideal flow is guaranteed by the convergence of the Markov chain. Since transportation network is always strongly connected network (otherwise one can go somewhere without the ability to go home), the stochastic matrix is always irreducible and based on Perron-Frobenius theorem, the maximal eigenvalue is one and therefore ideal flow algorithm is converged.
2. Through the use of the stochastic matrix to replace OD matrix, we gain several interesting properties such as scaling ideal flow matrix by a positive constant produces identical ideal flow and an ideal flow matrix is always premagic because of flow conservation on nodes.
3. Ideal flow may be a good indicator of the importance of the edge. Higher ideal flow may indicate possible relative higher flow level or congestion (if equal capacity). This edge relative importance is based on the network structure rather than the utilization of the network. Thus, it is based on inherent properties of the graph structure. This characteristics of dependency only on the network structure is very powerful concept because now can do what if analysis through simulation the network without worry of additional information of the actual demand.
4. We have at least three tools to check the structural relative importance of an edge: manual computation, linear algebra and multi-agent random walk simulation. The

    relative flow distribution of random walk on network is asymptotically equal to manually traced distribution of flow over the network with uniform distribution. Increasing the number of agents or maximum simulation time will create better flow distribution, closer to the actual value of ideal flow. Manual computation works only for small network. For larger network, we should use simulation and linear algebra approaches. Simulation methods works for large network but the result is only an approximation to the actual value of the ideal flow distribution. Linear algebra approach seems to be the best method so far.
5. Adding edge on network may increase or decrease congestion level due to increase of importance level of the edge. Ideal flow is a good tool to evaluate network dynamic that can give us the ability to see the effect of change in a network even far away from the place of change.
6. The proposed Ideal Flow method was demonstrated to real world network of Sioux Fall transportation network that has been used as standard test for traffic assignment. The results showed that with only a single parameter of scaling, the network flow distribution was fitted nearly perfect.

    An ideal traffic flow would be the target objective of network optimization. Having an ideal traffic flow will also lead researchers to discover indices to measure how far our current traffic situation the ideal condition. A change in a network structure could also change the ideal traffic flow correspond to the network. An ideal traffic flow also shows the limitation of current network. We can measure how far we can go with the current network structure and what can we do about it to improve the current situation. Having the ideal flow distribution, we found more open questions than what we can answer at this moment such as
1. What make additional edge lead to higher / lower ideal flow?
2. Is there a better way to compute generalized ideal flow using linear algebra for large network of a real city?
3. What other sample of network structure can reveal more properties? What other network properties should be investigated?
4. What are other properties of ideal flow matrix?
5. Increase of the number of agents or maximum simulation time does not change the distribution of the flow. Where these numbers of the flow distribution of the simulation coming from?
6. Since the relative flow distribution is more related to the network structure than utilization, what network properties can be correlated to the relative flow distribution?
7. It would be nice to check the ideal flow of the actual road. Does it represent the actual congestion level?
8. Is there any mathematical relationship between ideal flow and Braes Paradox?

    Some possible extensions for applications of ideal flow for vulnerability or reliability analysis would be suggested for further study.
    Many transport network modelling software such as CUBE, Transims, Aimsun, TransCAD has traffic assignment module and it is also the author wishes that such software would include Ideal Flow solution as part of the their module in the near future.

**ACKNOWLEDGEMENTS**
This research is supported by Commission of Higher Education (CHED) through Philippine Higher Education Research Network (PHERNET).

**LIST OF NOTATIONS**

| | |
|---|---|
| **A** | adjacency matrix of the directed graph |
| **B** | incidence matrix |
| $\mathbf{B}^-$ | negative values of the incidence matrix |
| $\mathbf{B}^+$ | positive values of the incidence matrix |
| **C** | constraint matrix |
| **D** | augmented matrix |
| $E$ | edges set |
| **e** | edge flow vector |
| **f** | flow ratio vector (normalized edge flow) |
| **F** | ideal flow matrix |
| $f_{ij}$ | flow on a link from node $i$ to node $j$ |
| G | directed graph |
| $H$ | network entropy |
| **j** | vector ones |
| $L$ | Lagrangian |
| **M** | premagic matrix |
| N | number of random walker agents |
| **n** | vector of node values |
| **R** | flow matrix based on count of agent trajectories |
| $\mathbf{s}_1^T$ | row vector sum of columns |
| $\mathbf{s}_2$ | column vector sum of rows |
| T | simulation time of random walk |
| **T** | transition probability matrix (Markov stochastic matrix) |
| $v_t$ | node at iteration t |
| $\Delta$ | diagonal matrix of each node degree |
| $\lambda$ | eigenvalue, or Lagrange multiplier (based on context) |
| $\kappa$ | scaling parameter |
| **π** | vector of stationary distribution of nodes (Perron's vector) |
| $\pi$ | scalar node value |